\documentclass[preprint,3p,twocolumn,12pt]{elsarticle}

\usepackage{graphicx}

\usepackage{amssymb}

\usepackage{amsmath}

\usepackage{booktabs} 

\usepackage{wasysym}

\usepackage{textcomp}

\usepackage{multirow}

\usepackage{aas_macros} 

\newcommand{\sub}[1]{_{\mathrm{#1}}} 

\journal{Physics Letters A}

\begin{document}

\begin{frontmatter}

\title{Damping parametric instabilities in future gravitational wave
  detectors by means of electrostatic actuators}

\author[mit]{John Miller\corref{cor1}\fnref{fn1}}
\ead{john.miller@anu.edu.au}

\author[mit]{Matthew Evans}
\author[mit]{Lisa Barsotti}
\author[mit]{Peter Fritschel}
\author[mit]{Myron MacInnis}
\author[mit]{Richard Mittleman}
\author[mit]{Brett Shapiro}
\author[mit]{Jonathan Soto}
\author[cit]{Calum Torrie}

\cortext[cor1]{Corresponding author}
\fntext[fn1]{Permanent address: Centre for Gravitational Physics, The
  Australian National University, Acton 0200, Australia. Tel: +61
  2 6125 1012, Fax: +61 2 6125 0741.}

\address[mit]{LIGO Laboratory, Massachusetts Institute of Technology, 185
  Albany St, Cambridge, MA 02139, United States of America}

\address[cit]{LIGO Laboratory, California Institute of Technology, 1200
   E. California Blvd., Pasadena, CA 91125, United States of America}

\begin{abstract}
  It has been suggested that the next generation of interferometric
  gravitational wave detectors may observe spontaneously excited
  parametric oscillatory instabilities. We present a method of
  actively suppressing any such instability through application of
  electrostatic forces to the interferometers' test masses. Using
  numerical methods we quantify the actuation force required to damp
  candidate instabilities and find that such forces are readily
  achievable. Our predictions are subsequently verified experimentally
  using prototype Advanced LIGO hardware, conclusively demonstrating
  the effectiveness of our approach.
\end{abstract}


\end{frontmatter}

\section{Introduction}
\label{sec:Introduction}
The second generation of interferometric gravitational wave detectors
\cite{Ward09,AdvancedVIRGOWebpage,Kuroda2010} shall utilise test
masses of high mechanical quality factor and operate with increased
circulating power in their Fabry-Perot arm cavities. These changes are
designed to reduce test mass thermal noise and quantum shot noise
respectively (see e.g.~\cite{Harry06,Caves81}). However, current
understanding suggests that these alterations will also make advanced
detectors more susceptible to parametric instabilities --
opto-acoustic interactions with the ability to excite mechanical
eigenmodes of an interferometer's optics, compromising detector noise
performance and control \cite{Braginsky01,Strigin07,Evans10,Gras10}.

A number of schemes are currently being investigated to mitigate this
potential problem \cite{AntlerThesis,Ju09}. Here we consider one such
scheme employing electrostatic actuators to actively damp any
instability in a frequency selective manner.

Using numerical methods we estimate the damping available from these
actuators and find that it is sufficient to quash all theoretically
dangerous modes. Our numerical results are subsequently verified
experimentally using prototype Advanced LIGO hardware.

Although we focus our attention on the Advanced LIGO detectors our techniques
are equally applicable to other long-baseline interferometers.

\subsection{Parametric instability}
\label{sec:PI}
The kilometre-scale Fabry-Perot arm cavities of second generation
interferometric gravitational wave detectors are expected to store
$\sim$1~MW of optical power. The mirrors forming these cavities
consist of multi-layer dielectric coatings deposited atop 40~kg
cylinders (34~cm~\diameter~$\times$~20~cm~tk.) of fused silica. These
cylinders are known as \emph{test masses} and exhibit mechanical
quality factors $\gtrsim10^7$.

Parametric instabilities (PIs) result from the non-linear coupling of
optical energy stored in an interferometer's arm cavities into
mechanical energy stored in the internal mechanical modes of its test
masses. This coupling is driven by ponderomotive radiation pressure
forces and may be easily understood as a classical feedback effect
\cite{Evans10}.

Excitation, thermal or otherwise, of a test mass eigenmode initiates
the process, scattering light from the fundamental cavity mode into
sideband fields with frequency spacing equal to the mechanical mode
frequency. The spatial profile of the test mass eigenmode is also
imprinted onto these fields. The sidebands then experience the optical
response of the interferometer, which is generally different for upper
and lower sidebands, before arriving back to the excited optic
together with the main cavity field. Radiation pressure couples energy
from the scattered light into mechanical motion, thus closing the
feedback loop. Based on overall loop phase, the mechanical mode may be
suppressed or further excited, the latter case being potentially
unstable. This process may be described by a single dimensionless
quantity, the parametric gain $R_m$ \cite{Evans10},
\begin{equation}
\label{eq:R}
R_m = \frac{4\pi
  Q_mP\sub{circ}}{M\omega_m^2c\lambda_0}\sum^\infty_{n=0}\Re [G_n]B^2_{m,n}.
\end{equation}
Here $Q_m$ is the quality factor of test mass mode $m$, $\omega_m$ is
its angular resonant frequency, $M$ is the mass of the test mass,
$P\sub{circ}$ is the circulating optical power in the main cavity mode
and $\lambda_0$ is its wavelength. $B_{m,n}$ is the geometric overlap
of mechanical mode $m$ with optical eigenmode $n$.  $\Re[G_n]$ is the
real part of the complex transfer coefficient from field $n$ leaving
an optic's surface to a field incident on and then reflected from the
same surface; for details of its calculation see \cite{Evans10} and
references therein.

\begin{figure}[htbp!]
\centering
\includegraphics[width=0.985\columnwidth]{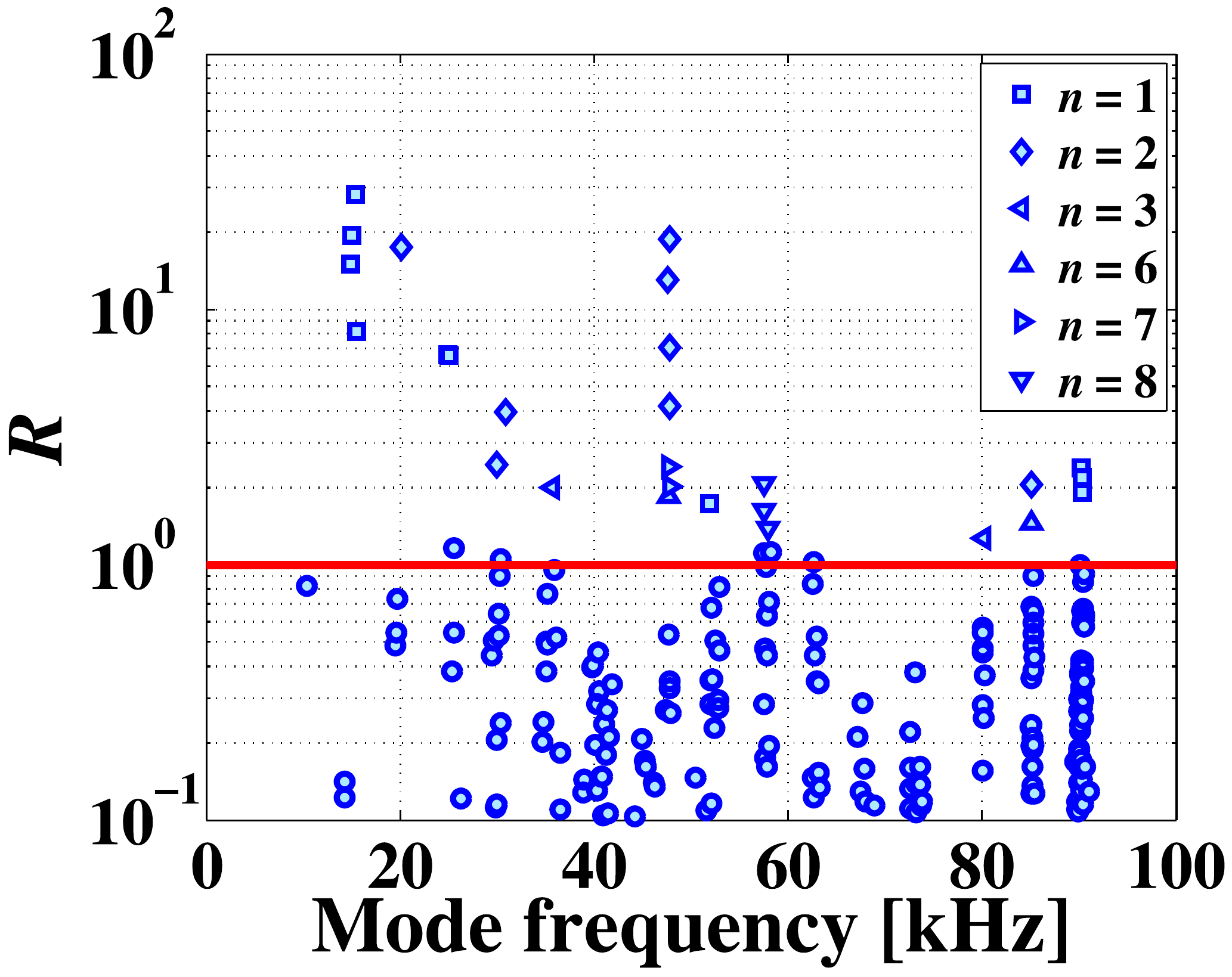}
\caption{``Worst case'' $R_m$ values for an interferometer
  configuration representative of Advanced LIGO. In our evaluation we
  randomly sweep over realistic model parameters, setting an upper
  limit on $R$ for each mode as the lowest value greater than 99.5\%
  of the results. Diffraction losses are included via the clipping
  approximation. We find 212 modes with $R>0.1$ and 32 modes with
  $R>1$. For selected instabilities, the order of the optical mode
  most strongly excited is given in the legend.}
\label{fig:Rvalue}
\end{figure}

Modelling the system as a feedback loop, $R_m$ is nothing other than
the real part of the open loop gain. Thus instability occurs for
$R_m>1$. The parametric gain has been evaluated for an Advanced LIGO
configuration (see table \ref{tab:RvalueParameters}), including stable
recycling cavities and diffraction loss, using the ``worst case''
methods of Evans et~al.~\cite{Evans10}. Results, shown in figure
\ref{fig:Rvalue}, reveal a number of potentially unstable modes.
\begin{table*}[htbp!]
  \caption{Interferometer parameters used in the
    theoretical evaluation of parametric gain. These values correspond
    to those of Advanced LIGO.}
\centering
\begin{tabular*}{\textwidth}{@{\extracolsep{\fill}}p{3cm}lc}
 \toprule
  &Quantity &Value \\
  \midrule
  \multirow{3}{3cm}{Lengths} &  Arm cavity& 3994.5 m\\
  &Power recycling cavity & 57.175 m\\
  &  Signal recycling cavity& 55.475 m\\
  \midrule
 \multirow{3}{3cm}{Gouy phases} &  Arm cavity & 156$^\circ$\\
  &  Power recycling cavity & 25$^\circ$\\
  &  Signal recycling cavity & 20$^\circ$\\
  \midrule
  \multirow{7}{3cm}{Optical properties}  &  Circulating arm power & 1 MW\\
  &  Input mirror power transmittance & 0.014\\
  &  End mirror power transmittance& $10^{-5}$\\
  &  Power recycling mirror power transmittance& 0.03\\
  &  Signal recycling mirror power transmittance& 0.2\\
  &  Beam splitter power transmittance& 0.5\\
  &  Laser wavelength& 1064 nm\\
  \midrule
  \multirow{2}{3cm}{Mechanical properties}  &  Mass of test mass & 40 kg\\
  &  Mechanical mode Q& $10^7$\\
  \bottomrule
\end{tabular*}
 \label{tab:RvalueParameters}
\end{table*}

As our analysis suggests, parametric instabilities may prove
troublesome in second generation interferometers. Hence, efforts are
ongoing to evaluate methods of ameliorating them. Schemes currently
under study fall into two broad categories: those which modify $G_n$
\cite{Degallaix07,Fan2010} and those which modify $Q_m$
\cite{AntlerThesis,Gras08}. In this work we investigate the
possibility of modifying the $Q$ of test mass mechanical modes using
an electrostatic actuator.

\subsection{Electrostatic drive}
\label{sec:ESD}
The majority of currently operating gravitational wave detectors
utilise coil-magnet actuators to control the position and angular
orientation of their suspended optics. The permanent magnets attached
to the interferometer test masses are known to introduce noise through
a number of mechanisms \cite{Acernese06,Yamamoto02,Manson72}. In order
to eliminate these problems, Advanced LIGO uses Electrostatic Drives
(ESDs) for the control of its most sensitive optics. Similar ESDs have
been used successfully in the GEO 600 interferometer for some time
\cite{Hewitson03}.

\begin{figure}[htbp!]
\centering
\includegraphics[width=\columnwidth]{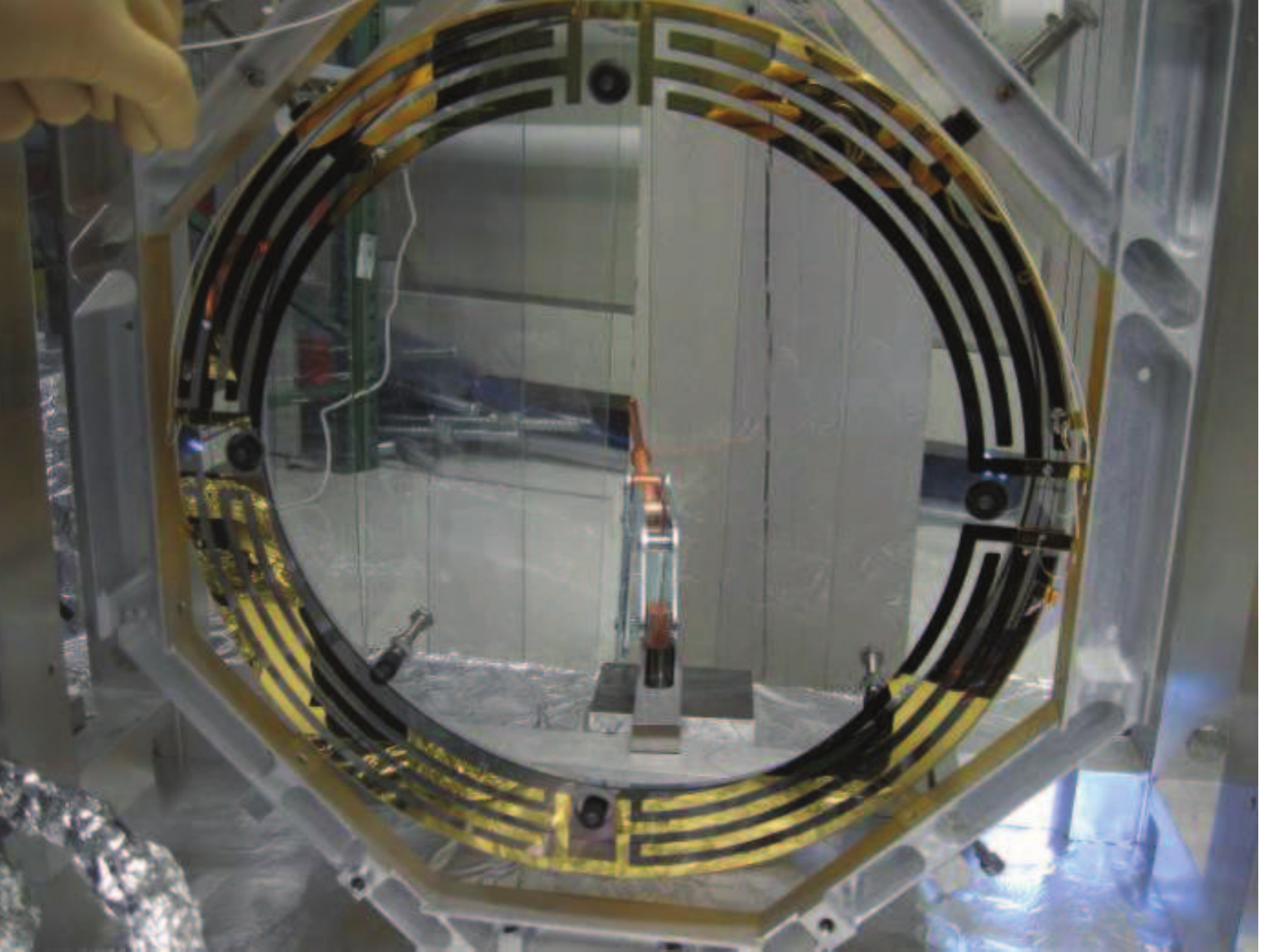}
\caption{Advanced LIGO noise prototype reaction mass and ESD. This
  pattern is used for all numerical and experimental work discussed
  herein.}
\label{fig:LASTIesd}
\end{figure}
\begin{figure}[htbp!]
\centering
\includegraphics[width=0.35\columnwidth]{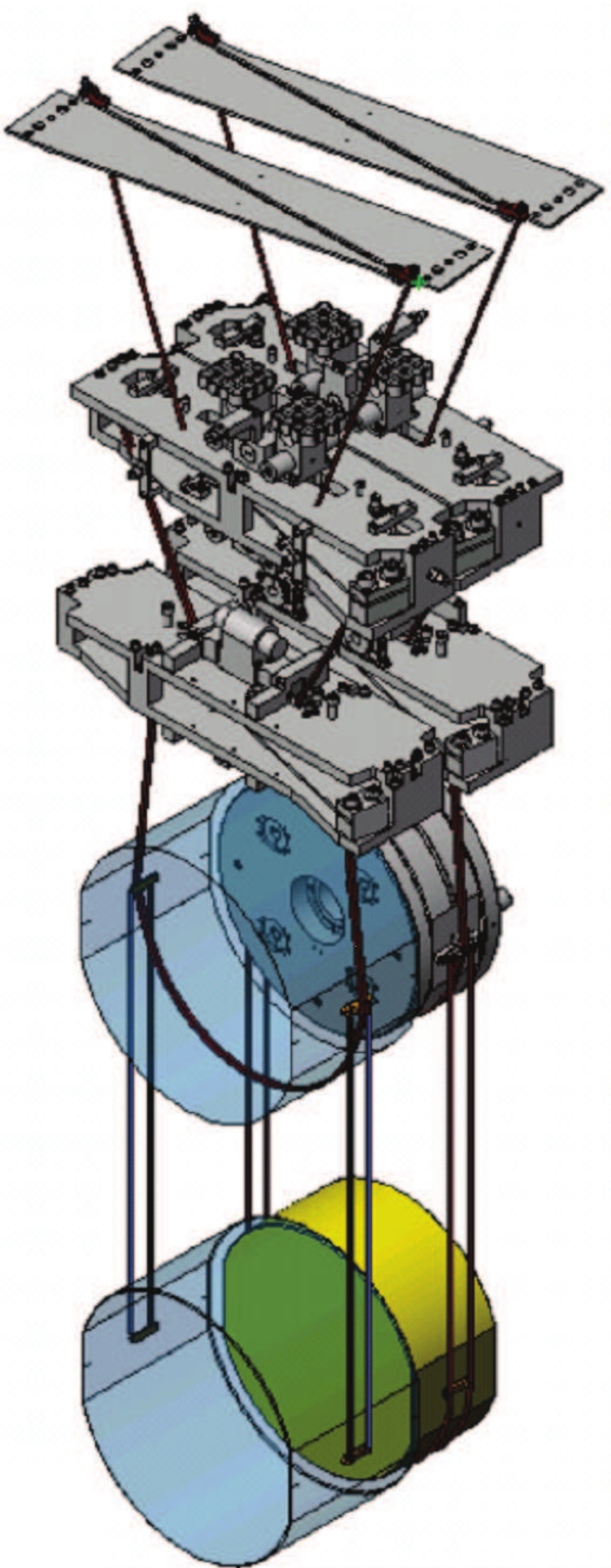}
\caption{Engineering rendering of Advanced LIGO quadruple suspension
  system. The upper pendulum stages will continue to use coil-magnet
  actuators with only the critical final stage meriting an ESD.  The
  ESD pattern is deposited onto the surface of the reaction mass
  (highlighted in yellow) adjacent to the test mass. Figure
  \ref{fig:LASTIesd} shows a prototype reaction mass in more detail.}
\label{fig:Quad}
\end{figure}

For concreteness and in order to facilitate verification by
experiment, we limit our discussion to the particular ESD shown in
figure \ref{fig:LASTIesd}. This ESD, an Advanced LIGO prototype
considered for control of the input test masses, consists of a
symmetric pattern of four electrode pairs. These electrodes are
deposited in gold on the surface of a fused silica \emph{reaction
  mass}. The reaction mass is positioned immediately behind the test
mass with the ESD electrodes facing its rear surface, so that the
electrostatic force is applied between the two masses. The nominal
inter-mass spacing is 5~mm. To attenuate seismic disturbances both the
reaction mass and test mass are suspended from isolated platforms by
quadruple pendulums (see figure \ref{fig:Quad}).

A potential difference between the ESD's electrode pairs gives rise to
fringing electric fields which attract the dielectric test mass. This
attractive force is proportional to the square of the potential
difference and may be characterised by
\begin{equation}
\label{eq:alpha}
F\sub{ESD}=\alpha(\Delta V)^2,
\end{equation}
where $\Delta V$ is the potential difference between the electrode
pairs and $\alpha$ is a constant of proportionality dependent on the
separation between the test mass and reaction mass, the material
properties of the two optics and the specific geometry of the ESD
pattern. With four separate electrode pairs, both angular orientation
and longitudinal position can be controlled.

\section{Modelling}
\subsection{Required force}
In order for the ESD to effectively damp any parametric instability it
must have a discernible effect on the $Q$ of test mass mechanical
modes. Using arguments outlined fully in \ref{sec:ForceCalculation} we
find that, for a thermally excited mode, the magnitude of force
demanded from the ESD to reduce the parametric gain from $R_m$ to
$R_{\mathrm{eff},m}$ is
\begin{equation}
\label{eq:Fesd}
 F_{\mathrm{ESD},m}=\sqrt{\mu_mk_BT}\frac{\omega_{0,m}}{b_m}\bigg(\frac{R_m-R_{\mathrm{eff},m}}{Q_mR_{\mathrm{eff},m}}\bigg).
\end{equation}
Here $\omega_{0,m}$ is the angular eigenfrequency of mode $m$, $\mu_m$
is its modal mass and $Q_m$ its mechanical quality factor. $T$ is the
ambient temperature, taken to be 300~K, and $k_B$ is Boltzmann's
constant. The parameter $b_m$, defined via (\ref{eq:ESDoverlap}),
represents the fraction of applied ESD force which is coupled into
mechanical mode $m$. Given the difficulties involved in evaluating
this parameter analytically, a numerical approach was employed.

\subsection{Evaluating $b_m$}
In order to calculate $b_m$ one requires knowledge of both the test
mass eigenmodes and the spatial distribution of ESD forces.

Mode shapes and frequencies of an Advanced LIGO test mass, including
the flat regions on the barrel of the mass\footnote{The inclusion of
  flat regions splits modes without axial symmetry into doublet pairs
  \cite{Strigin2008}.} but not the `ears' which link the mass to its
suspension fibres, were found using commercial finite element software
\cite{COMSOL,ANSYS}. This analysis produced several thousand
eigenmodes with frequencies in the \hbox{10-90 kHz} band. We limit
ourselves to this region as higher frequency modes typically couple to
high order optical modes which suffer from significant diffraction
losses and thus yield low parametric gain. In our investigation only
modes with appreciable parametric gain ($R_m>0.1$) were considered.

As a consequence of the complex electrode geometry and non-uniform
electric fields involved, the calculation of the force produced by the
ESD is also well suited to finite element analysis. An appropriate
model was constructed and solved for the electrostatic
potential. Selected model parameters are presented in
table~\ref{tab:FEMParameters}.
\begin{table}[htbp!]
  \caption{Parameters used in modelling the performance of the
    electrostatic drive.}
\centering
\begin{tabular*}{\columnwidth}{@{\extracolsep{\fill}}lc}
 \toprule
  Quantity &Value \\
  \midrule
  Test mass radius  & 17~cm\\
  Test mass thickness & 20~cm\\
  Reaction mass radius  & 17~cm\\
  Reaction mass thickness & 13~cm\\
  Inter-mass spacing & 5~mm\\
  \midrule
 Relative permittivity of masses & 3.75\\
 Inner radius of electrode pattern & 13.3~cm\\
  Outer radius of electrode pattern & 16.8~cm\\
  Electrode thickness & 5~mm\\
  Electrode spacing & 5~mm\\
  \bottomrule
 \end{tabular*}
\label{tab:FEMParameters}
\end{table}

\begin{figure}[htbp!]
\centering
\includegraphics[width=\columnwidth]{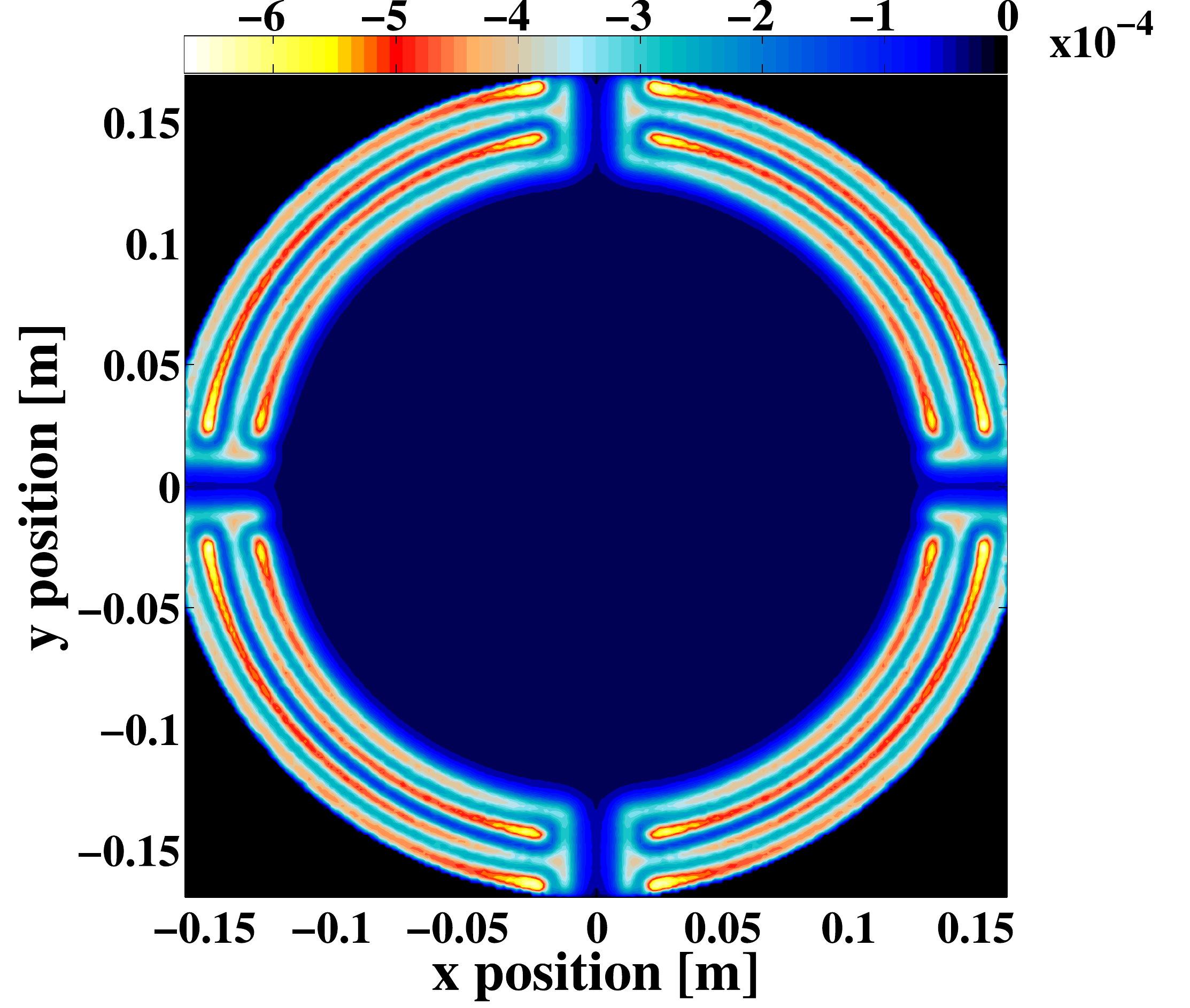}
\caption{ESD pressure distribution in Pa as a function of transverse
  position across the test mass. Forces are uniformly attractive
  (negative) and most significant in the vicinity of the electrode
  pattern. $\Delta V=200$ V is assumed.}
\label{fig:ForceDensity}
\end{figure}

With a view to finding the ESD pressure distribution $p\sub{ESD}$
across the transverse dimensions of the test mass, the Maxwell stress
tensor (MST) $\Upsilon_{ij}$ was invoked. For our electrostatic
simulation
\begin{equation}
\Upsilon_{ij} = E_iD_j-\frac{1}{2}\delta_{ij}\sum^3_{k=1}E_kD_k, 
\end{equation}
where $E_i$ and $D_i$ are the $i^{\mathrm{th}}$ Cartesian components
of electric and displacement fields and $\delta_{ij}$ is the Kronecker
delta. Integrating the MST over discrete closed volumes within the
test mass, the pressure was obtained as a function of position (see
figure~\ref{fig:ForceDensity}). For further detail regarding this
calculation see \cite{MillerThesis}.

With mode shapes and pressure distribution available, the theoretical
$b_m$ force coupling coefficients were found (via
(\ref{eq:ESDoverlap})) and used to evaluate the force required to damp
parametric instabilities using an electrostatic drive.

\subsection{Modelling results}
A critical parameter for lock acquisition and interferometer control
is the peak force available from the ESD. Integrating $p\sub{ESD}$
over its domain\footnote{Or equivalently integrating the MST over the
  test mass boundary.} we find the maximum available force to be
$\sim$190~\textmu N for \hbox{$\Delta V = 800$~V}, a reasonable value
for a wide bandwidth, low-noise voltage amplifier and one considered
for Advanced LIGO. This force corresponds to \hbox{$\alpha = 2.9\times
  10^{-10}\mbox{ N/V}^2$}.

\begin{figure}[htbp!]
\centering
\includegraphics[width=\columnwidth]{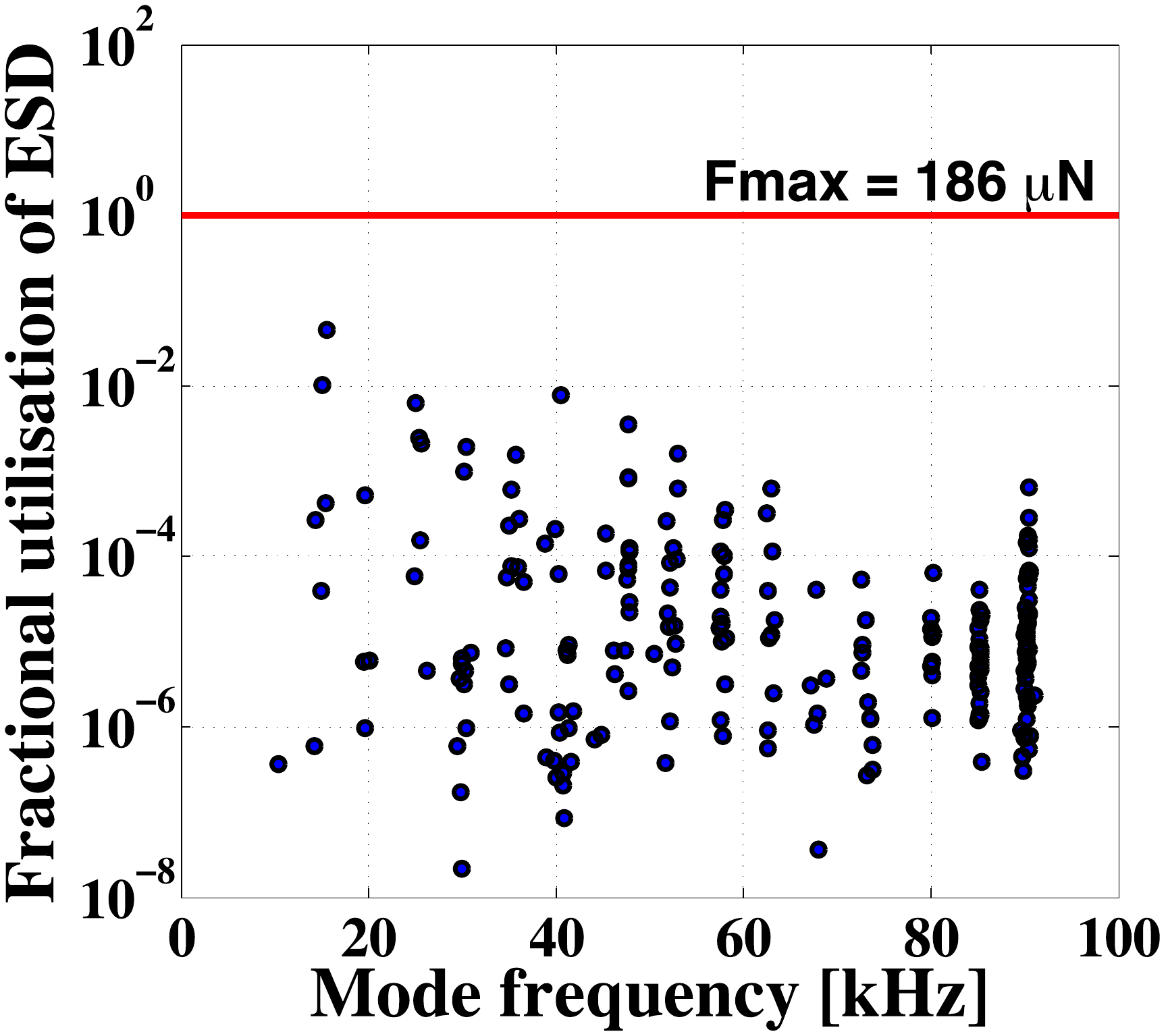}
\caption{Fraction of total force available from all four ESD quadrants
  required to reduce the parametric gain to 0.1. Maximum utilisation
  is 4.6\%. Available force evaluated for $\Delta V=800$ V.}
\label{fig:ForceRequiredFullPattern}
\end{figure}

Figure \ref{fig:ForceRequiredFullPattern} shows the fraction of this
total force which must be drawn upon to reduce the effective
parametric gain of each mode to 0.1 or less, a tolerable value for
long-baseline interferometers. In all cases we see that successful
damping is theoretically achievable using just a few percent of the
available force.

\begin{figure}[htbp!]
\centering
\includegraphics[width=\columnwidth]{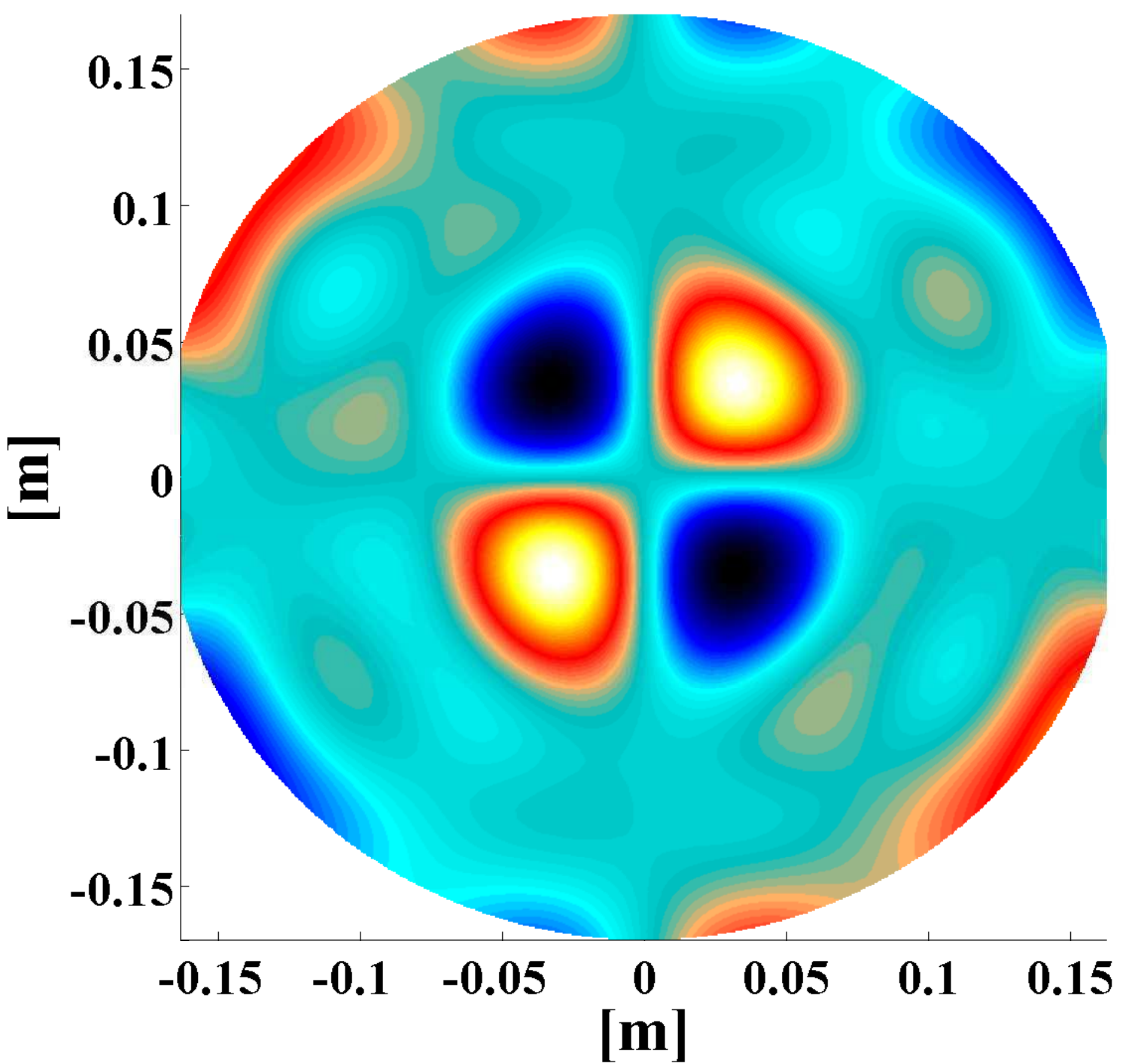}
\caption{Mechanical modes with low order symmetries could prove
  difficult to damp using the four-fold symmetric ESD. This problem
  can be mitigated by introducing small asymmetries to the ESD
  pattern, by operating with just a single ESD quadrant or by
  recruiting all four ESD quadrants with mode dependent phasing.}
\label{fig:SymmetricMode}
\end{figure}

Due to diffraction loss, optical gain is low for higher order
transverse modes. This suggests that lower order optical modes,
excited by low order test mass eigenmodes, are most likely to give
rise to PIs. These low order mechanical modes exhibit low order
symmetries, implying that their coupling to the four-fold symmetric
ESD pattern will be poor and that they will thus require large damping
forces (see figure \ref{fig:SymmetricMode}).

To combat this effect asymmetric ESD patterns were explored. It was
quickly recognised that large benefits were available by operating
with just a single ESD quadrant. Whilst median values of the damping
force are comparable in both cases, the maximum utilisation of
available force is reduced by more than an order of magnitude. This
leads us to recommend single quadrant operation as the baseline mode
for our scheme.
\begin{figure}[htbp!]
\centering
\includegraphics[width=\columnwidth]{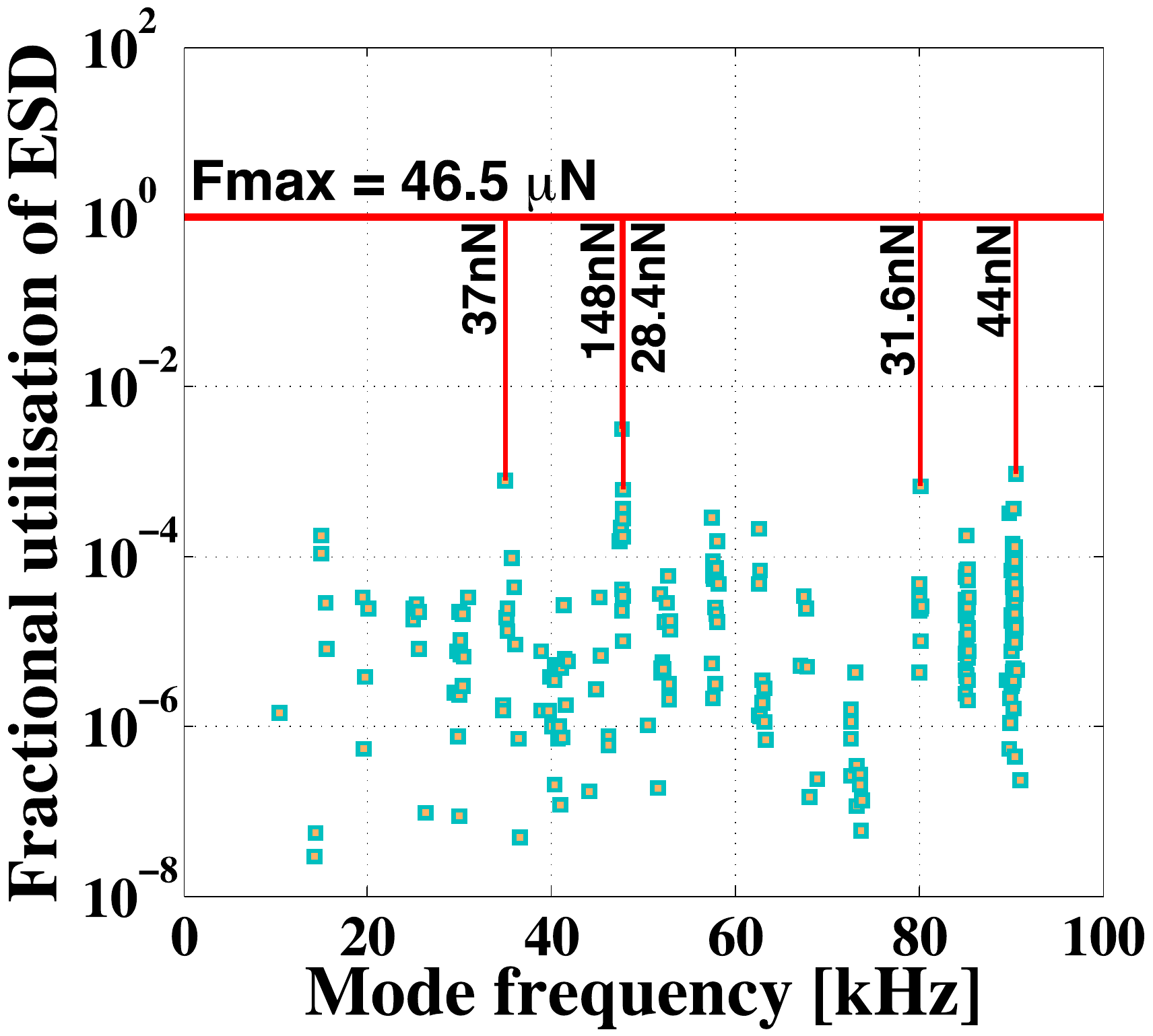}
\caption{Fractional utilisation of available force to realise $R=0.1$
  in single quadrant operation. Maximum utilisation is 0.3\%. The
  absolute force required to damp each of the five most demanding
  modes is indicated. Again, $\Delta V=800$~V is assumed.}
\label{fig:ForceRequiredOneQuad}
\end{figure}

Operating in this way will also allow ESD damping to remain effective
even if the test mass-reaction mass separation is increased from the
nominal value of 5~mm.\footnote{The test mass-reaction mass separation
  at the ITMs may be increased by a factor of $\sim$4 to
  reduce gas damping noise \cite{Cavalleri2010}. Although the
  available force will fall by factor of $\sim$40 effective damping is
  still realisable.}

The fraction of available force utilised in single quadrant operation
is shown in figure \ref{fig:ForceRequiredOneQuad}. Should additional
actuation be required, the various quadrants can be recruited
simultaneously with mode dependent phasing.

\section{Experimental verification}
\subsection{Apparatus}
In order to confirm the predictions of our numerical model an
experimental investigation was carried out. All measurements were made
by studying induced length fluctuations of an independently suspended,
digitally controlled, 16~m Fabry-Perot optical resonator.

Our work utilised several Advanced LIGO prototype systems. Most notably, the
cavity end mirror was a prototype 40~kg Advanced LIGO test mass, suspended by
the quadruple pendulum system shown in figure~\ref{fig:Quad} and
controlled by the ESD shown in figure~\ref{fig:LASTIesd}. The cavity
input mirror was not used for prototyping test mass assemblies and
thus had only a triple suspension and no reaction mass.

The resonator was driven by a sub-W Nd:YAG NPRO, frequency stabilised
to a rigid reference cavity. Control of the suspended cavity was
achieved using standard techniques \cite{Drever83} with control
signals being applied to the coil-magnet actuators of the triple
suspension. The unity gain frequency of the suspended cavity servo was
set to 100 Hz.

With this apparatus we were able to measure $\alpha$ and investigate
our theoretically predicted couplings, $b_m$.

\subsection{The ESD actuation coefficient $\alpha$}
Our numerical prediction for $\alpha$ was tested experimentally by
driving the cavity length at low frequencies (10-100 Hz) using all
four quadrants of the ESD and measuring the response in the cavity
error signal. To maximise available force, the potential of all
electrodes was set to the largest possible value, with the two
electrodes in each pair having opposite polarity.

For excitation frequencies above the highest pendulum resonance and
below the first internal mode of the test mass we may write the
peak-to-peak ESD force as
\begin{equation*}
F_{\mathrm{pp}}=M\omega_{\mathrm{exc}}^2x_{\mathrm{pp}},
\end{equation*}
where $M$ is again the mass of the test mass, $\omega_{\mathrm{exc}}$
is our excitation frequency in rads$^{-1}$ and $x_{\mathrm{pp}}$ is
the peak-to-peak test mass displacement along the cavity
direction.

For known excitation waveforms, the peak-to-peak
differential voltage $\Delta V_{\mathrm{pp}}$ applied to the ESD is
also known. Thus from (\ref{eq:alpha}) we have
\begin{equation*}
\alpha =\frac{F_{\mathrm{pp}}}{\Delta V_{\mathrm{pp}}^2}.
\end{equation*}

\begin{figure*}[htbp!]
\centering
\includegraphics[width=0.8\textwidth]{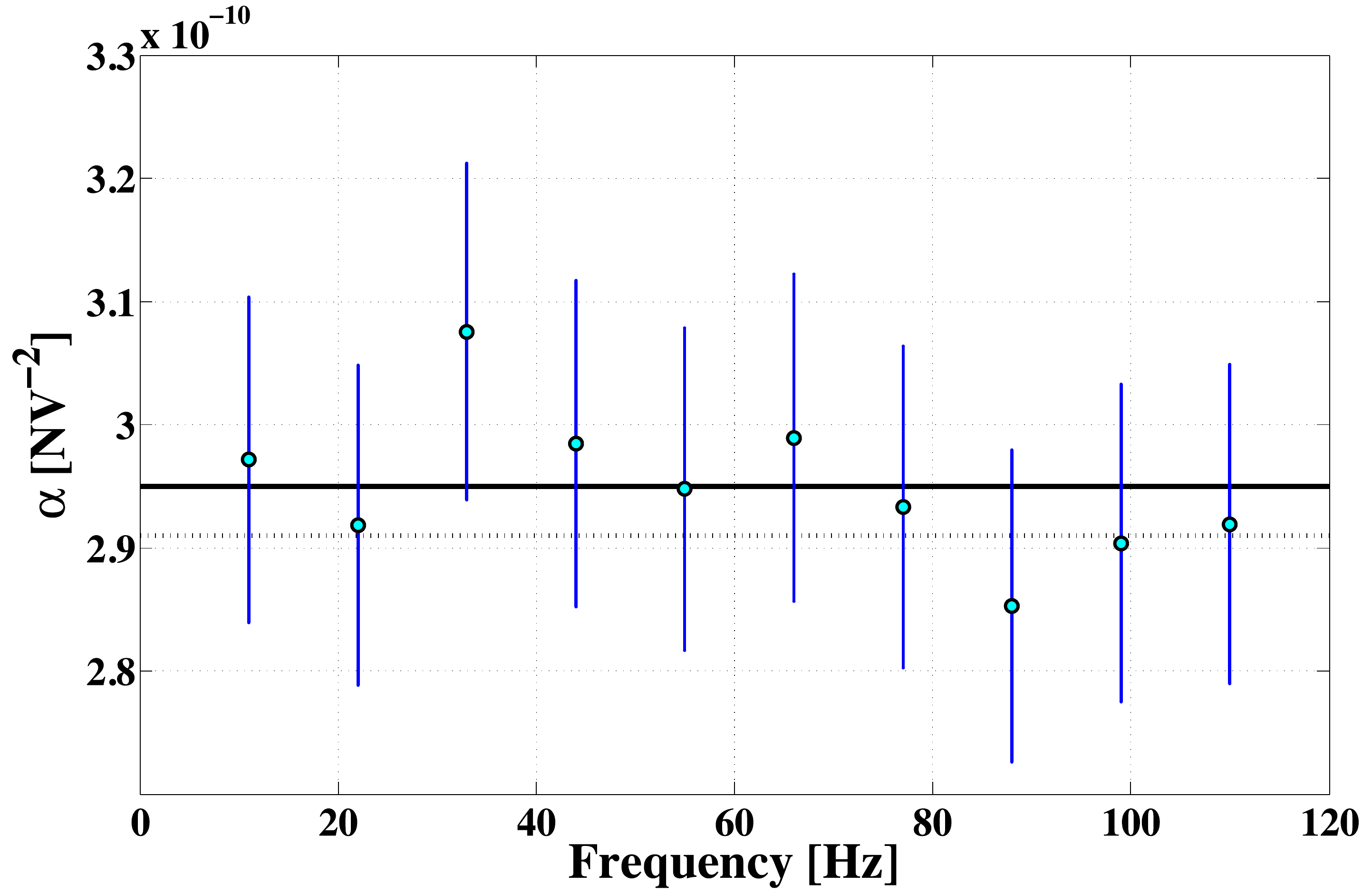}
\caption{ESD actuation coefficient measurement. The solid line
  indicates the mean value of \hbox{$2.95\times 10^{-10}\mbox{ N/V}^2$}; the
  dashed line shows the theoretically predicted value of $2.91\times
  10^{-10}\mbox{ N/V}^2$.}
\label{fig:alpha}
\end{figure*}

Measurements of $x\sub{pp}$ were made as a function of excitation
amplitude and frequency; $\alpha$ was found to have an mean value of
\hbox{$2.95\pm0.08\times 10^{-10}\mbox{ N/V}^2$} (see figure
\ref{fig:alpha}). Accord with the predicted value of \hbox{$2.91\times
  10^{-10}\mbox{ N/V}^2$} is excellent.

\subsection{The ESD-eigenmode coupling coefficients $b_m$}
The force coupling between the ESD and test mass mechanical modes is
of key importance in accurately quantifying our capacity to damp
instabilities. In order to validate the theoretically calculated
values, a series of measurements was made.

A simple feedback loop was implemented to excite the mechanical mode
of interest into steady state oscillation. Breaking the loop allowed
ringdown measurements to be made.

Knowing the steady state modal amplitude achieved for a given drive
waveform and having measured the $Q$ via ringdowns, the ESD-mechanical
mode overlap was found by modelling the system as a simple oscillator
with angular resonant frequency $\omega_0$. At equilibrium, we have
for each mode
\begin{equation}
b_m = \frac{\omega_{0,m}^2\mu_m
  A_m}{Q_mF_{\mathrm{ESD},m}},
\label{eq:bm}
\end{equation}
where $\mu_m$ is the modal mass of mode $m$.

The amplitude of $F_{\mathrm{ESD},m}$ is calculated as
\begin{equation*}
F_{\mathrm{ESD},m} = \frac{1}{2}\alpha \bigg[V_{\mathrm{exc},m}-V\sub{bias}\bigg]^2,
\end{equation*}
where $V\sub{exc}$ is the analogue\footnote{The internal modes of the
  test mass lay beyond the Nyquist frequency ($\sim$8 kHz) of our
  digital system.} excitation voltage applied to the inner ESD
electrode and $V\sub{bias}$ the constant bias voltage applied to the
outer electrode.

The modal amplitude $A_m$ is determined from the cavity error signal
via \hbox{$A_m=x_m/c_m$}, where $x_m$ is the measured test mass
displacement along the cavity direction and $c_m$ is a geometric
overlap factor which accounts for the fraction of mechanical mode
amplitude which is sensed by the cavity.\footnote{For example, if the
  cavity beam is incident on the mirror near a node $c$ will be
  smaller than if the beam were incident near an anti-node.} This
parameter is calculated as
\begin{equation*}
c_m = \bigg|\iint\limits_{\mathcal{S}}I_{00}(\vec{u}_m\cdot\hat{z})
\,\mathrm{d}\mathcal{S}\bigg|.
\end{equation*}
Here $\vec{u}_m$ has the same normalisation as before (see
(\ref{eq:UMnormalisation})) and $I_{00}$, the intensity profile of the
cavity's principal optical mode, has an integrated power of 1 W.

In evaluating $c_m$ we make two assumptions. Firstly, that we know the
shape of the mode which has been excited and secondly, that the cavity
optical mode is centred on the test mass.

\begin{table}[htbp!]
\caption{Experimental mode parameters.}
\centering
\begin{tabular}{cccccc}
\toprule
f measured &f predicted &$Q$                 \\
kHz        &kHz         &                    \\
\midrule
8.149      &8.1439      & 412,170 $\pm$ 1003 \\
10.4115    &10.397      & 604,158 $\pm$ 4728\\
12.9705    &12.959      & 508,666 $\pm$ 2473\\
15.0405    &15.0439     & 409,504 $\pm$ 4139\\
\bottomrule
\end{tabular}
\label{tab:OverlapResults}
 \end{table}
The close agreement between the mode frequencies predicted by finite
element modelling and those observed experimentally (see table
\ref{tab:OverlapResults}) gives us confidence that the correct mode
shapes were considered.

\begin{figure}[htbp!]
\centering
\includegraphics[width=0.49\columnwidth]{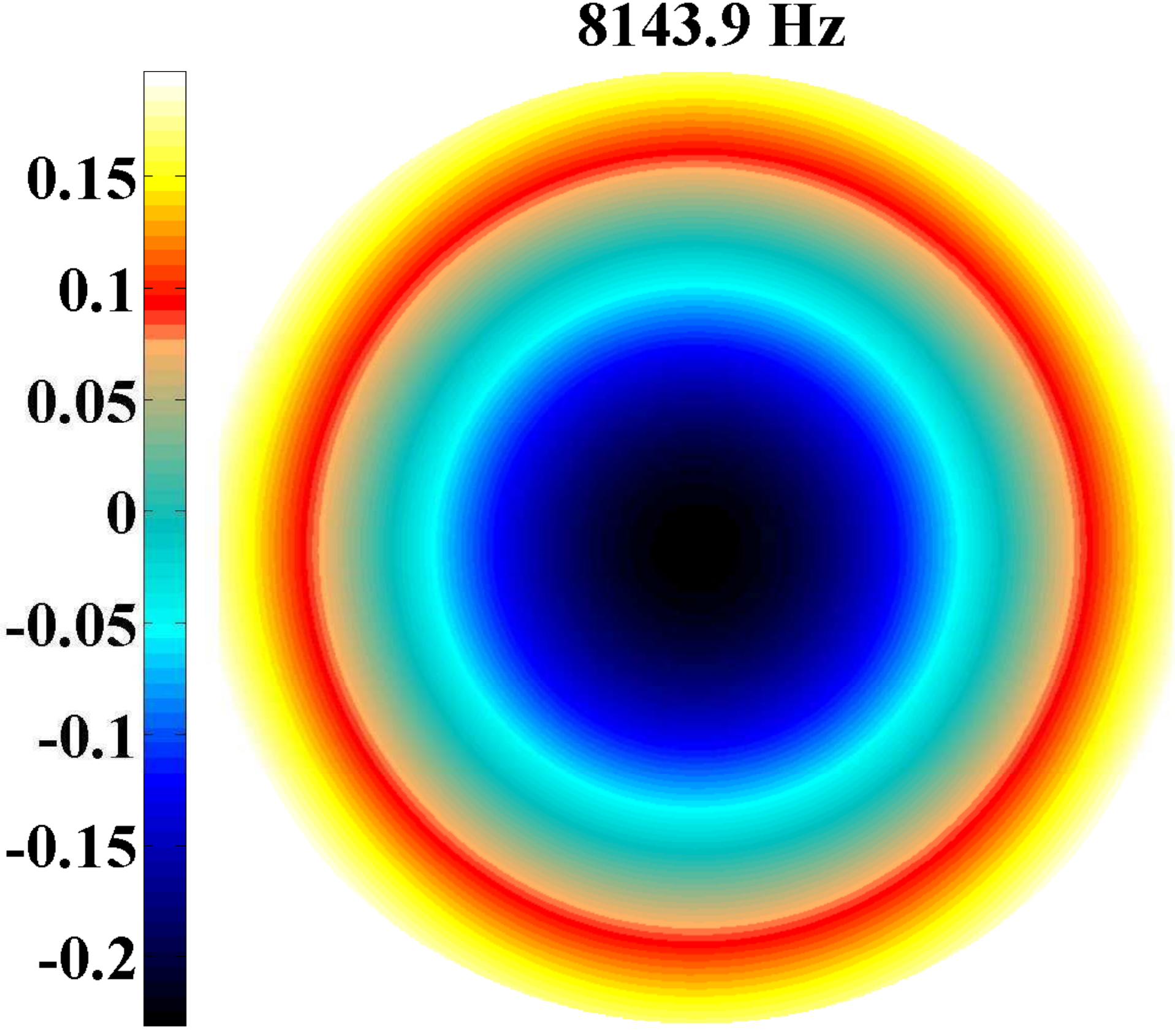}
\includegraphics[width=0.48\columnwidth]{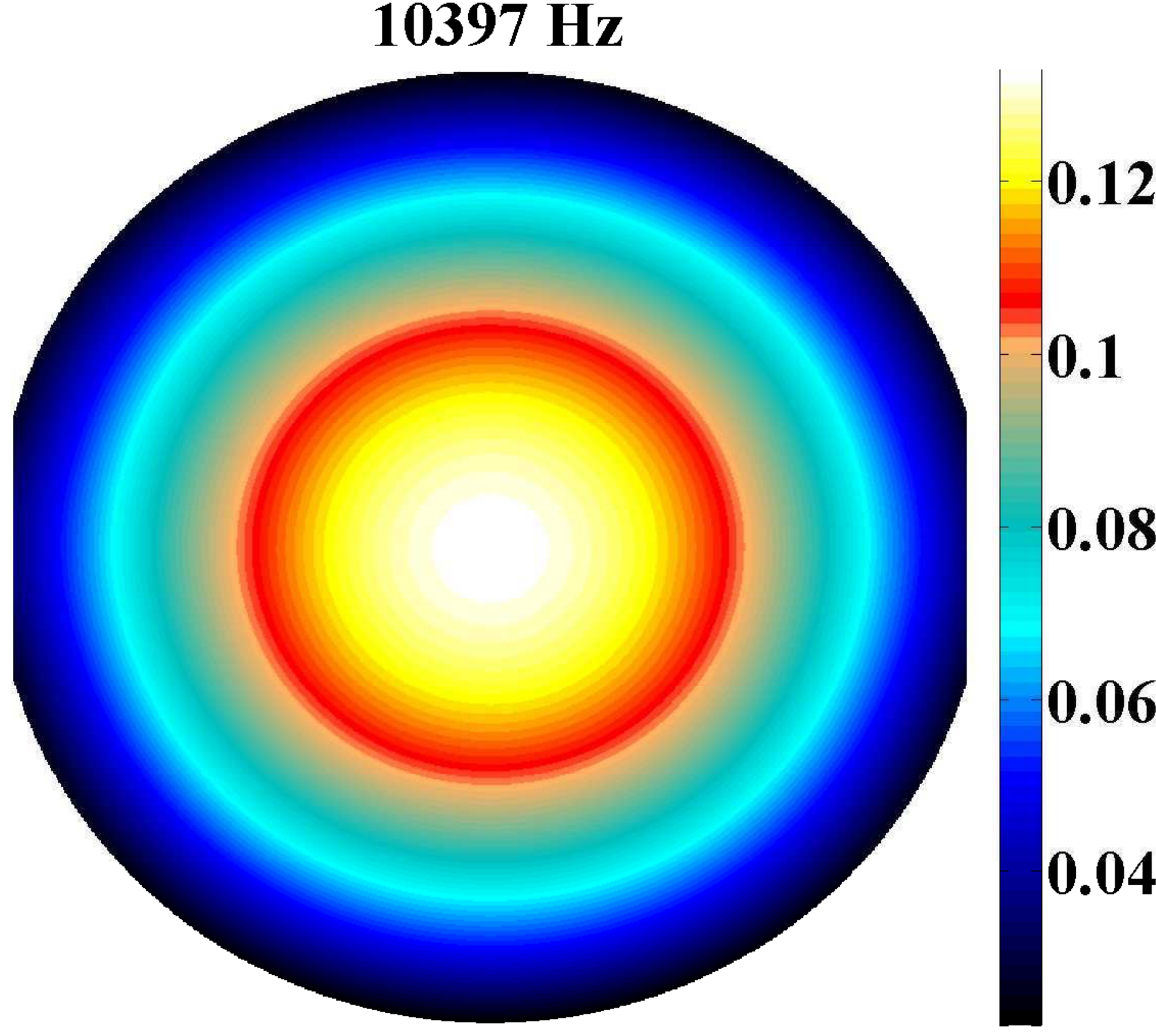}\\
\includegraphics[width=0.49\columnwidth]{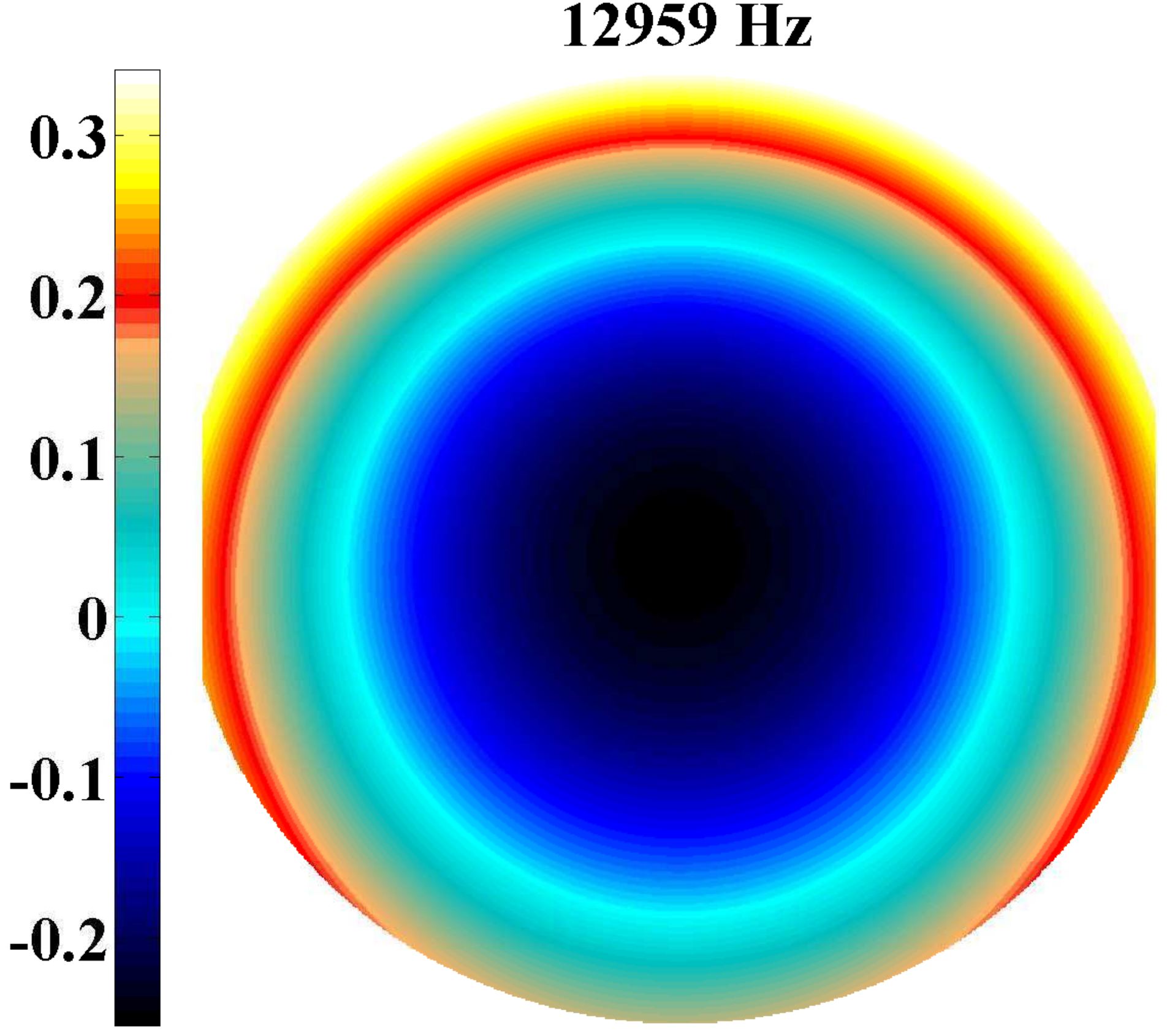}
\includegraphics[width=0.48\columnwidth]{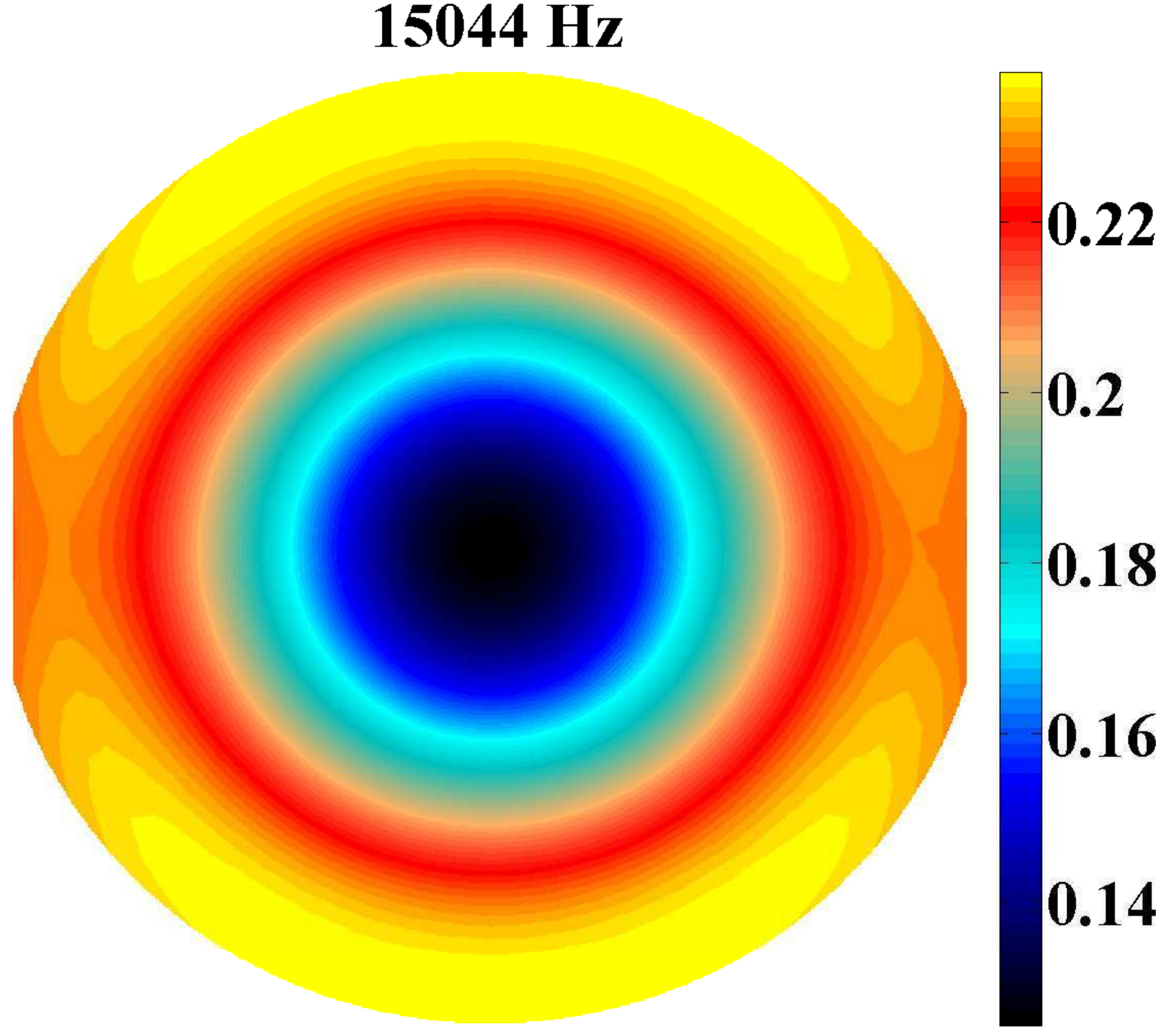}
\caption{Modelled face displacements, $\vec{u}_m\cdot\hat{z}$, of the
  experimentally studied modes. The asymmetry in the 13 kHz mode is
  due to the wedge angle of the test mass.}
\label{fig:modes}
\end{figure}

To mitigate any errors introduced by deviations from perfect centring
we limited our analysis to `drumhead' type modes with smoothly varying
displacements near their centres. The four modes considered are shown
in figure \ref{fig:modes}.

Experimentally measured $Q$ values, listed in table
\ref{tab:OverlapResults}, were found to be significantly lower than
the value of $10^7$ we assume in our theoretical calculations. Note,
however, that these $Q$s were measured with a metal wire suspension;
Advanced LIGO will adopt a quasi-monolithic fused silica design
\cite{Robertson02}. The test mass discussed herein has recently been
suspended from silica fibres. Initial measurements confirm that $Q$s
of order 10$^7$ will be achievable in second generation
interferometers.

\begin{figure}[htbp!]
\centering
\includegraphics[width=\columnwidth]{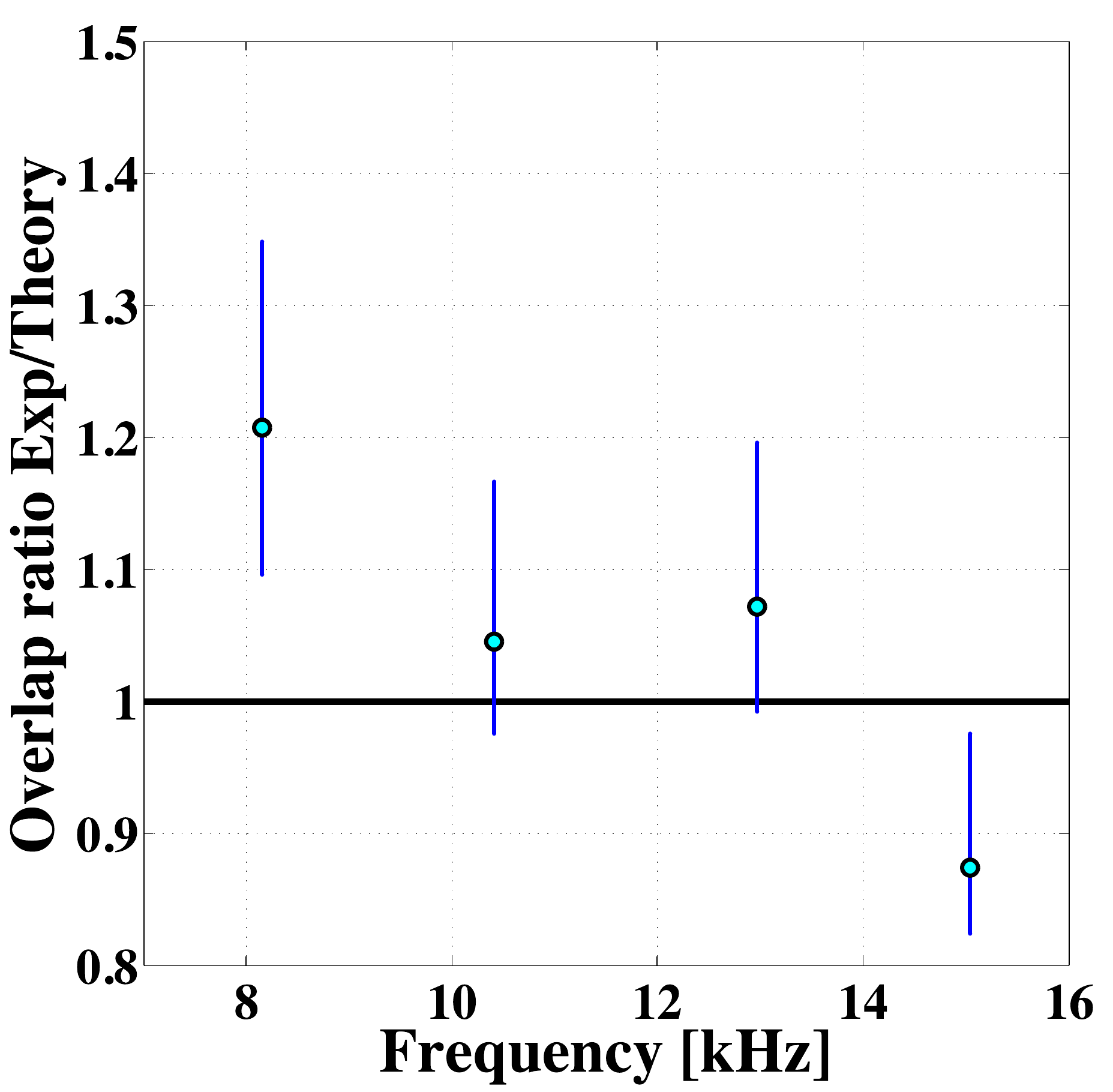}
\caption{Ratio of experimental to theoretical overlap coefficients
  $b_m$ for the test mass modes of figure \ref{fig:modes}.}
\label{fig:Overlaps}
\end{figure}

Using the apposite experimentally measured information, (\ref{eq:bm})
was evaluated to yield $b_m$. Comparing our results to theoretical
predictions, agreement was found to be better than 20\%. The ratio of
experimental to theoretical overlap coefficients is plotted in figure
\ref{fig:Overlaps}.

\section{Summary and conclusions}
By numerical methods we have obtained the eigenmodes of a prototype
Advanced LIGO test mass and the pressure distribution available from
the electrostatic actuator used in its control. Having experimentally
verified both quantities, this information was used to predict the
force required to damp parametric instabilities in Advanced LIGO. We find that
such forces are readily available. This result clearly demonstrates
that electrostatic actuators do indeed represent a viable means of
damping parametric instabilities.

In contrast with other techniques \cite{AntlerThesis,Gras08}, our
damping scheme will not introduce any additional thermal
noise. Further, due to the frequency selective nature of our approach,
we also anticipate that any sensing noise injected by the ESD will be
negligible. Unless a test mass has an appropriately placed eigenmode
it will not exhibit any appreciable response to our damping
forces.\footnote{This fact also allows us to apply damping forces to
  all test masses with impunity.}  What little coupling there is will
be evident outside of the gravitational wave detection band.

There will of course be practical challenges to using electrostatic
damping to control parametric instabilities. For example, it is
possible for a number of modes to be above the instability threshold
simultaneously. Although a feature of ESD damping is its ability to
target each mode individually, this would also necessitate
implementing a discrete control loop for each mode, increasing
complexity as the number of problematic modes increased.

Each instability must also be sensed through the presence of the
higher order optical modes which it excites. Extraction of these
signals is complicated by the coupled resonant cavities found in
advanced interferometers.

If PIs really do begin to threaten the robust operation of second
generation interferometers there will likely be no single solution. It
is probable that a combination of approaches will be employed, perhaps
based around passive damping with an active scheme like ours targeted
to particularly troublesome modes. Electrostatic actuators have proven
themselves to be effective in suppressing test mass $Q$s and are
therefore capable of playing a major role in any such mitigation
scheme.

\appendix
\section{Force required to damp mechanical modes}
\label{sec:ForceCalculation}

Considering each acoustical mode of the test mass as a damped
oscillator with resonant frequency $\omega_{0,m}$ rads$^{-1}$ and
modal mass $\mu_m$ we have
\begin{equation}
\label{eq:oscillator}
\mu_m\ddot{x}_m+\frac{\omega_{0,m}\mu_m}{Q_m}\dot{x}_m+k_mx_m-F_{\mathrm{app},m}=0,
\end{equation}
where $F_{\mathrm{app},m}$ is the force applied by the ESD. Assuming
viscous damping
\begin{equation*}
F_{\mathrm{app},m}=-K_m\dot{x}_m,
\end{equation*}
we write
\begin{equation*}
\mu_m\ddot{x}_m+\bigg[\frac{\omega_{0,m}\mu_m}{Q_{\mathrm{eff},m}}\bigg]\dot{x}_m+k_mx_m=0,
\end{equation*}
in doing so defining an effective $Q$,
\begin{equation*}
Q_{\mathrm{eff},m}=\bigg(\frac{1}{Q_m}+\frac{K_m}{\mu_m\omega_{0,m}}\bigg)^{-1},
\end{equation*}
whence
\begin{equation*}
K_m=\mu_m\omega_{0,m}\bigg(\frac{1}{Q_{\mathrm{eff},m}}-\frac{1}{Q_m}\bigg).
\end{equation*}
Therefore, asserting that $\dot{x}_m=j\omega_{0,m}x_m$,
\begin{equation*}
F_{\mathrm{app},m} =j\mu_m\omega_{0,m}^2x_m\bigg(\frac{1}{Q_m}-\frac{1}{Q_{\mathrm{eff},m}}\bigg).
\end{equation*}
For thermal excitations, \hbox{$x_m=(k_BT/\mu_m\omega_{0,m}^2)^{1/2}$}, we
have
\begin{equation*}
F_{\mathrm{app},m}= j\omega_{0,m}\sqrt{\mu_mk_BT}\bigg(\frac{1}{Q_m}-\frac{1}{Q_{\mathrm{eff},m}}\bigg).
\end{equation*}

We now introduce an additional parameter $b_m$, defined through
\hbox{$F_{\mathrm{app},m}=b_mF_{\mathrm{ESD},m}$}, to represent the coupling
between the ESD actuation force and the mechanical mode in
question. $b_m$ may be calculated as follows,
\begin{equation}
\label{eq:ESDoverlap}
b_m=\bigg|\iint\limits_{\mathcal{S}}p_{\mathrm{ESD}}(\vec{u}_m\cdot\hat{z})\,\mathrm{d}\mathcal{S}\bigg|,
\end{equation}
where $p_{\mathrm{ESD}}$ is the ESD pressure distribution and
$\vec{u}_m\cdot\hat{z}$ is the displacement of the mirror's surface
along the cavity axis. These quantities have normalisations
\begin{equation}
\iint\limits_{\mathcal{S}}p_{\mathrm{ESD}}\,\mathrm{d}\mathcal{S}=1\mbox{
and }\iiint\limits_{\mathcal{V}}\rho
|\vec{u}_m|^2\,\mathrm{d}\mathcal{V}=1,
\label{eq:UMnormalisation}
\end{equation}
$\rho$ being the uniform mass density of the test mass, $\mathcal{S}$
the rear surface of the optic normal to the cavity axis and
$\mathcal{V}$ the test mass volume. With this normalisation $\mu_m =1$
for all modes.

Thus, including the overlap parameter $b_m$ and making use of the
proportionality of $Q$ and $R$, i.e.
\begin{equation*}
  Q_{\mathrm{eff},m} = Q_m\frac{R_{\mathrm{eff},m}}{R_m},
\end{equation*}
the magnitude of the required ESD force is given by
\begin{equation*}
\boxed{
\begin{gathered}
F_{\mathrm{ESD},m}=
\sqrt{\mu_mk_BT}\frac{\omega_{0,m}}{b_m}\bigg(\frac{R_m-R_{\mathrm{eff},m}}{Q_mR_{\mathrm{eff},m}}\bigg).
\end{gathered}
}
\end{equation*}

\section*{Acknowledgements}
\label{sec:Acknowledgements}
The authors would like to thank Kenneth Strain, Norna Robertson, James
Hough and Rainer Weiss for their valuable input to this project. JM's
work at The Massachusetts Institute of Technology was carried out
under the auspices of the LIGO Visitors Program and supported by the
Carnegie Trust for the Universities of Scotland.  LIGO was constructed
by the California Institute of Technology and the Massachusetts
Institute of Technology with funding from the National Science
Foundation and operates under cooperative agreement PHY-0757058. This
paper has LIGO Document Number LIGO-P1000122.

\bibliographystyle{model1a-num-names}
\bibliography{MillerBibliography}

\end{document}